\documentclass{aa}  
\usepackage{color}
\usepackage{float}
\usepackage{lineno}
	\usepackage{caption}

\usepackage{graphicx}
\usepackage{txfonts}
\usepackage{subcaption}         
\usepackage{lscape}             
\usepackage{placeins}           
                                
\begin{document}

\title{More power on large scales}

\author{Jeremy Mould$^{1,2}$}
\institute{Swinburne University, PO Box 218, Hawthorn 3121, Australia\\
\and ARC Centre of Excellence for Dark Matter Particle Physics\\
             \email{jmould@swin.edu.au}
}
   \date{Received November 30, 2025}

  \abstract
  {	
  The high value
	of the cosmic microwave dipole may be telling us that dark matter is macroscopic
	rather than a fundamental particle. 
	The possible presence of a significant dark matter component in the form of primordial black holes (PBHs) sug
gests that dark halo formation simulations should be commenced well before redshift z = 100. 
	Unlike standard cold dark matter candidates, which are initially relativistic or possess thermal velocities, 
PBHs behave as dense, non-relativistic matter from their inception in the radiation-dominated era. This allows them to seed gravitational potential wells and begin clustering much earlier, significantly altering the initial power spectrum on small scales. We find that starting N-body simulations at redshifts even before matter-radiation equality (z 
$\sim$ 3400) yield galaxy bulk flow velocities that are systematically larger than those predicted by standard $\Lambda$CDM models.
	The early, high-mass concentrations established by PBHs lead to a more rapid and efficient gravitational acceleration of surrounding baryonic and dark matter, generating larger peculiar velocities that remain coherent over scales of hundreds of Mpc.
	Furthermore, a sub-population of PBHs in the 10$^{-20}$ to 10$^{-17}$ M$_\odot$ mass range would lose a non-negligible fraction of their mass via Hawking radiation over cosmological timescales. This evaporation process convert
s matter into radiation, so  a time-varying matter density parameter, $\Omega_m^\prime$, is introduced, which behaves
 like a boosted radiation term, 
	 in the Friedmann equation. This dynamic term, which is most active between recombination and the late universe, acts to reduce the Hubble tension. A higher effective $\Omega_r$ in the early universe (pre-evaporation) reduces 
the sound horizon at the epoch of recombination. This smaller "standard ruler," as imprinted on the cosmic microwave 
background (CMB), would result in a higher  
 value of the Hubble constant (H$_0$) 
	inferred from the CMB 
at the 1\% level, bringing it into slightly closer agreement with local, late-time measurements. 
	PBH mass loss also influences fits to the equation of state
	parameter, w, at low redshift. The naive N-body  modelling presented here suggests investigation with
	tried and tested cosmology codes should be carried out, by introducing mass losing PBHs and starting
	the evolution as early as practicable.}
\keywords{Primordial black holes(1292) -- Cosmology(343) -- Large-scale
structure of the universe(902) }

   \maketitle


\title{More power on large scales}
\author{Jeremy Mould$^{1,2}$}
\institute{Swinburne University, PO Box 218, Hawthorn 3121, Australia
\and ARC Centre of Excellence for Dark Matter Particle Physics}


\abstract
{	{\color{red} The high value
	of the cosmic microwave dipole may be telling us that dark matter is macroscopic
	rather than a fundamental particle.}
	The possible presence of a significant dark matter component in the form of primordial black holes (PBHs) suggests that dark halo formation simulations should be commenced well before redshift z = 100. 
	Unlike standard cold dark matter candidates, which are initially relativistic or possess thermal velocities, PBHs behave as dense, non-relativistic matter from their inception in the radiation-dominated era. This allows them to seed gravitational potential wells and begin clustering much earlier, significantly altering the initial power spectrum on small scales. We find that starting N-body simulations at redshifts even before matter-radiation equality (z $\sim$ 3400) yield galaxy bulk flow velocities that are systematically larger than those predicted by standard $\Lambda$CDM models.
	The early, high-mass concentrations established by PBHs lead to a more rapid and efficient gravitational acceleration of surrounding baryonic and dark matter, generating larger peculiar velocities that remain coherent over scales of hundreds of Mpc.
	Furthermore, a sub-population of PBHs in the 10$^{-20}$ to 10$^{-17}$ M$_\odot$ mass range would lose a non-negligible fraction of their mass via Hawking radiation over cosmological timescales. This evaporation process converts matter into radiation, so  a time-varying matter density parameter, $\Omega_m^\prime$, is introduced, which behaves like a boosted radiation term, 
	 in the Friedmann equation. This dynamic term, which is most active between recombination and the late universe, acts to reduce the Hubble tension. A higher effective $\Omega_r$ in the early universe (pre-evaporation) reduces the sound horizon at the epoch of recombination. This smaller "standard ruler," as imprinted on the cosmic microwave background (CMB), would result in a higher  
 value of the Hubble constant (H$_0$) 
	inferred from the CMB 
at the 1\% level, bringing it into slightly closer agreement with local, late-time measurements. 
	PBH mass loss also influences fits to the equation of state
	parameter, w, at low redshift. The naive N-body  modelling presented here suggests investigation with
	tried and tested cosmology codes should be carried out, by introducing mass losing PBHs and starting
	the evolution as early as practicable.}
\keywords{Primordial black holes(1292) -- Cosmology(343) -- Large-scale
structure of the universe(902) }
\maketitle

\end{document}